\documentclass[11pt]{article}
\usepackage{graphicx}
\usepackage{amsmath,amssymb,amsfonts}
\newcommand{\comment}[1]{{}}

\title{A note on quantization of matrix models}

\author{Artem Starodubtsev\thanks{email: astarodu@astro.uwaterloo.ca} \\
 \\ \centerline{\footnotesize \it Department of Physics, University of Waterloo} \\ \centerline{\footnotesize \it 200
University ave W, Waterloo, ON, Canada N2L 3G1} \\
\centerline{\footnotesize \it and} \\
\centerline{\footnotesize \it Perimeter Institute for
Theoretical Physics} \\
\centerline{\footnotesize \it 35 King st N, Waterloo, ON, Canada
N2J 2W9 }}

\date{}

\begin{document}

\maketitle

\begin{abstract}
The issue of non-perturbative background independent quantization
of matrix models is addressed. The analysis is carried out by
considering a simple matrix model which is a matrix extension of
ordinary mechanics reduced to 0 dimension.   It is shown that this
model has an ordinary mechanical system  evolving in time as a
classical solution. But in this treatment the action principle
admits a natural modification which results in algebraic relations
describing quantum theory. The origin of quantization is similar
to that in Adler's generalized quantum dynamics. The problem with
extension of this formalism to many degrees of freedom is solved
by packing all the degrees of freedom into a single matrix. The
possibility to apply this scheme to field theory and to various
matrix models is discussed.
\end{abstract}

\section{Introduction}
Matrix models were proposed as a non-perturbative definition of
string theory some time ago. There are several versions of matrix
models. Generally they are defined as reductions of certain Super
Yang Mills theories to one dimension (BFSS matrix model
\cite{BFSS}) or to zero dimension (IKKT matrix model \cite{IKKT}).

One of the most interesting properties of these models is the
dynamical origin of spacetime. For example the action of IKKT
model
\begin{equation}
S=-Tr ( \frac{1}{4} [A_\mu,A_\nu][A^\mu,A^\nu] + fermions )
\label{sikkt}
\end{equation}
does not contain any a priori spacetime structure at all. The
later arises from classical solutions of the model (\ref{sikkt})
and it is associate to the distribution of the eigenvalues of the
matrices $A_\mu$ \cite{stiib}. Spacetime is said to be "generated
dynamically" in these models.

Besides spacetime structure matrix models also give rise to
various objects propagating in spacetime such as D-instantons,
D-strings, D0-branes, membranes etc., depending on a particular
model chosen, and also to local interactions between those
objects. This provides some evidence for the conjecture that
matrix models can be a constructive non-perturbative definition of
string theory.

Because the definition (\ref{sikkt}) of a matrix model do not
include any background spacetime structure one may hope that this
model could provide a background independent definition of string
theory so that different string theories in various background
spacetimes would be different solutions of a single theory.

Classicaly the theory indeed looks background
independent\footnote[1]{The action (\ref{sikkt}) contains a fixed
flat Minkowskian metrics $\eta^{\mu \nu}$ contracting lower
indices, though. However in \cite{stiib} was considered a
possibility that this model can also in principle describe curved
spacetime. The metric $\eta^{\mu \nu}$ is to be understood in this
case as a metric in tangent space and the frame field forming the
metrics of the manifold originates from matrices. All the
background information encoded in $\eta^{\mu \nu}$ is the
dimension and the signature. Here it is worth mentioning also
cubic matrix models \cite{cmm} which do not depend on any
background metric structure at all.}. However the way it is
generally quantized is the following. First one picks up a
classical solution to the matrix model, $A_\mu=X_\mu$,
representing a particular spacetime structure with some background
fields in it. Then one considers small perturbations around the
classical solution
\begin{equation}
A_\mu=X_\mu+\tilde A_\mu
\end{equation}
representing the objects moving in the background spacetime and
interactions between them. Then one quantizes $\tilde A_\mu$. It
is clear that the resulting theory is no longer background
independent.

The key question which is addressed in this paper is whether we
can define a non-perturbative background independent quantization
of matrix models. The definition of non-prturbative quantization
of a theory is generally straightforward. One has to represent all
the observables and all the symmetries of the theory by operators
on a certain Hilbert space. However when we try to apply this to a
matrix model a certain puzzle emerges. Basic observables, matrices
$A_\mu$, which are to be represented by hermitian operators are
already hermitian operators on a certain Hilbert space. And the
basic symmetries of the theory, $U$ such that $S[U^{-1}A_\mu
U]=S[A_\mu]$, which are to be represented by unitary
transformations are already unitary operators in the same Hilbert
space. The straightforward application of the quantization
procedure then would lead us to something like "operator-valued
operators" which would be simply operators because any two Hilbert
spaces are isomorphic to each other. Therefore quantization
doesn't seem to change the shape of the theory.

A natural question then arises: Can matrix models produce quantum
theory without quantization in the usual sense? If so, how can it
be realized?

One proposal for this was made by Smolin in a recent paper
\cite{hiddenvar} which is that matrix model can be interpreted as
a non-local hidden variables theory, the eigenvalues being quantum
observables and the entries being hidden variables. Quantum
mechanics for the eigenvalues is then reproduced in the ordinary
statistical mechanical description of the model.

In this paper we will study the possibility that in matrix model
framework where there is no a priory spacetime structure classical
and quantum theory can be brought much closer to each other than
they are in the presence of a background spacetime. The later
would imply a possible relation between origin of space-time and
quantization. There are some indications on existence of such a
relation also pointed out in \cite{hiddenvar}. Many of notions on
the basis of which the distinction between classical and quantum
theory is generally made rely on the existence of classical
spacetime. For example classical system is recognized as having a
definite smooth trajectory, something what quantum system doesn't
have. But the very notion of smooth trajectory can not be defined
without using the the classical notion of spacetime and its
differential structure. Also quantum theory is characterized by
its generic non-locality. But the notion of locality also relies
on classical spacetime.  Given all the above it is very plausible
that matrix models with the generalized notion of spasetime that
they give rise to can provide framework which is general for
classical and quantum theory. All the distinction s between them
should be quantitative such as spectra of physical observables.

In the present paper the problem of quantization of matrix models
is studied on a simple example which is a matrix extension of
mechanical system. In section \ref{sec2} the model is defined. It
is shown that this matrix model reproduces ordinary mechanics. In
section \ref{sec3} a natural modification of this model is
proposed. The equations of motion of this modified model describe
quantum mechanics. Two alternative derivations of basic equations
of quantum mechanics are presented each giving the same result. In
section \ref{sec4} other possible modifications of the model are
discussed. They provide some generalizations of quantum mechanics
such as that with minimum length uncertainty relation. In section
\ref{sec5} the scheme is generalized to systems with many degrees
of freedom. This is done by a specific way of encoding the degrees
of freedom which is very natural for matrix models. In section
\ref{sec6} a possibility to generalize the model  to field theory
as well as to various matrix models related to string theory and
loop quantum gravity is discussed.

\section{Matrix model for a mechanical system with one degree of
freedom} \label{sec2} In this section we will define a matrix
model which reproduces the ordinary mechanics of a system with one
degree of freedom $x$ described by the following action  $S_{L}$
in the Lagrangian form.
\begin{equation}
S_{L}[x] =\int dt \Big\{ \frac{1}{2}\Big(\frac{dx}{dt} \Big)^2 -
V(x) \Big\} \label{s0}
\end{equation}
Here $V$ is an arbitrary function of $x$ which represents the
self-interaction of $x$. The equation of motion can be obtained by
requiring that the variation of  the action (\ref{s0}) with
respect to $x$ vanish.
\begin{equation}
\frac{\delta S_{L}}{\delta X(t)}=0. \label{vp0}
\end{equation}
This results in the standard Newtonian equation of motion:
\begin{equation}
\frac{d^2x}{dt^2}+\partial_x V(x)=0. \label{ne}
\end{equation}

All the constructions of this section can be equally done for both
the Lagrangian action principle (\ref{s0},\ref{vp0}) and the
Hamiltonian action principle
\begin{equation}
S_{H}[x,p] =\int dt \Big\{ p\frac{dx}{dt} -\frac{p^2}{2} - V(x)
\Big\}, \label{sH0}
\end{equation}
which is equivalent to (\ref{s0}) in the sense that  if minimize
this action with respect to the momentum $p$,
\begin{equation}
\frac{\delta S_{H}[x,p]}{\delta p}=0 \label{minp}
\end{equation}
and substitute the solution to the equation (\ref{minp}) $p=p(x)$
into the action (\ref{sH0}) we will find that
\begin{equation}
S_{H}[x,p(x)]=S_L[x],
\end{equation}
and therefore the action (\ref{s0}) and the action (\ref{sH0})
result in the same equation for $x$.  Below we will consider the
Lagrangian form of the matrix action. The generalization to the
Hamiltonian form is straightforward.

The basic entries of the action (\ref{s0}) are the configuration
coordinate $x$ and the time derivative $\frac{d}{dt}$ and also the
integration operation  $\int dt$. All this operations can be
thought of as realizations of a multiplication operator, a
derivative operator, and a trace operation respectively on a
certain space of functions $\psi(t)$.

\begin{eqnarray}
x(t)\times \psi(t) \ \ \ \leftrightarrow \ \ \ X \psi\nonumber \\
\frac{d}{dt} \psi(t) \ \ \ \leftrightarrow \ \ \ D \psi\nonumber \\
\int dt x(t) \ \ \ \leftrightarrow \ \ \ Tr X. \label{corresp}
\end{eqnarray}

 We do not specify the class of functions
$\psi(t)$ nor do we endow the space of them with any Hilbert space
structure. So all that matters is the algebraic relations between
those operators.

Given the correspondence (\ref{corresp})
 the action (\ref{s0}) can be rewritten in the
following operator form:
\begin{equation}
S_{L}[x]=S_{op}[d/dt,x(t)],
\end{equation}
where
\begin{equation}
S_{op}[D,X]=Tr \Big\{ \frac{1}{2}[D,X]^2 - V(X)    \Big\},
\label{opf}
\end{equation}
where $[A,B]=AB-BA$ is the operator commutator.

Now one can loosen  the requirement of the correspondence
(\ref{corresp}) and consider $D$ and $X$ to be unknown operator
valued variables. Of course the physical meaning of the operators
$X$ and $D$ as the position of the particle and the evolution
generator respectively has to be retained. All we get rid of is
the relation between them as between a smooth function of a
certain parameter and a derivative operator with respect to this
parameter.

Another way to define the action (\ref{opf}) which is more similar
to the usual way how matrix models are introduced is to consider
the action for a Higgs field in 0+1 dimension
\begin{equation}
S=\int d x^0 Tr[(D_0x)^2-V(x)]. \label{higgs}
\end{equation}
Here $x$ is the $SU(N)$ Higgs field and $D_0x=\partial_0x+[A_0,x]$
is the $SU(N)$ covariant derivative along $x^0$ direction. Now by
reduction of this model down to  0 dimension and taking the limit
$N \rightarrow \infty$ we obtain (\ref{opf}).

 Now the statement is that the matrix model defined by the action (\ref{opf})
 has the mechanical system defined by the action (\ref{s0}) as its solution.

To see this consider the variation of the action (\ref{opf}) with
respect to $D$ and $X$:
\begin{eqnarray}
\delta S_{op}[D,X]= Tr \bigg\{ \Big[ -[D,[D,X]] - \partial_X V(X)
\Big] \delta X \bigg\}+ Tr \Big\{  [D,\delta X[D,X]] \Big\}
\nonumber \\
+Tr \Big\{ [X,[D,X]]\delta D  \Big\}+ Tr \Big\{ [X,\delta D[D,X]]
\Big\}. \label{vari}
\end{eqnarray}
The second and the fourth terms in the variation would be equal to
zero if we were allowed to do cyclic permutations in the trace of
a product of several operators. The later is true for finite
dimensional operators but this property generally does not extend
to infinite dimensional ones. Here we will simply require that the
variations $\delta X$ and $\delta D$ are such that
\begin{equation}
Tr \Big\{  [D,\delta X[D,X]] \Big\}=0 \label{bc1}
\end{equation}
and
\begin{equation}
Tr \Big\{ [X,\delta D[D,X]] \Big\}=0. \label{bc2}
\end{equation}
The condition (\ref{bc1}) have a simple classical interpretation.
According to classical correspondence (\ref{corresp})we can
rewrite (\ref{bc1}) as
\begin{eqnarray}
0=Tr \Big\{  [D,\delta X[D,X]] \Big\} =\int\limits_{t_1}^{t_2} dt
\frac{d}{dt} \Big( \delta x(t) \frac{d x(t)}{dt} \Big) = \delta
x(t) \frac{d x(t)}{dt} \Big\vert_{t=t_1}^{t=t_2} \label{bc1cl}
\end{eqnarray}
The later holds when
\begin{equation}
\delta x(t_1)=\delta x(t_2)=0.
\end{equation}
Thus, the condition (\ref{bc1}) means that we vary the action with
respect to a function keeping it fixed at the endpoints. This is
how it is usually done in the Euler-Lagrange variational
principle. The condition (\ref{bc2}) has no classical analog and
therefore it is difficult to visualize.

As usual the variational principle is
\begin{equation}
\delta S_{op}[D,X]=0.
\end{equation}
Given the conditions (\ref{bc1}) and (\ref{bc2}) and that the
variations $\delta X$ and $\delta D$ are independent we have the
following equations of motion
\begin{equation}
[D,[D,X]] + \partial_X V(X)=0, \label{em1}
\end{equation}
\begin{equation}
[X,[D,X]]=0. \label{em2}
\end{equation}
Classically eq. (\ref{em1}) coincides with the newtonian equation
of motion. Eq. (\ref{em2}) has no classical analogs.

From now on we will look for solutions of operator equations such
as (\ref{em1},\ref{em2}) in a representation where $X$ is a
multiplication operator. Thus we assume that the operator algebra
can be realized on a space of functions $\psi(X)$. This means that
any operator from our algebra can be represented by the following
series (finite or infinite)
\begin{equation}
O=\sum\limits_{i,j=0}^{\infty}o_{ij}X^i \left(
\frac{d}{dX}\right)^j=
\sum\limits_{j=0}^{\infty}o_{j}(X)\left(\frac{d}{dX}\right)^j.
\label{patt}
\end{equation}
We can now use this form for $D$ and substitute it into
eq.(\ref{em2}). This will result in the following solution for $D$
\begin{equation}
D=d_1(X)\frac{d}{dX}+d_0(X), \label{solfod}
\end{equation}
where $d_1(X)$ and $d_2(X)$ are arbitrary functions of $X$. Now we
can  introduce a new variable $t$ such that
\begin{equation}
\frac{dx(t)}{dt}=d_1(x(t)). \label{proof5}
\end{equation}
Then  the solution to the equation (\ref{em2}) can be rewritten as
\begin{equation}
X=x(t)\times, \ \ \ D= \frac{d}{dt} +d_0(t). \label{prooff}
\end{equation}
Here $t$ is an arbitrary parameter, which means that the relations
(\ref{corresp}) were derived up to reparameterization. But if we
recall that physically by $D$ we mean the evolution generator in a
background time this freedom is fixed. Thus, all the relations
(\ref{corresp}) are recovered  and if we substitute them to the
initial action (\ref{opf}) we recover the classical action
(\ref{s0}). Therefore the matrix model based on the action
(\ref{opf}) describes classical dynamics of the system defined by
the action (\ref{s0}).

\section{Quantization without quantization} \label{sec3}

Now one can show that by a minor modification of the action
(\ref{opf}) the equations of motion (\ref{em1},\ref{em2}) can be
turned into those of quantum mechanics. The equations of motion in
this framework  are derived from variation of the action not only
with respect to $X$ but also with respect to $D$. Therefore the
action (\ref{opf}) admits a nontrivial modification by adding an
extra term to the action which depends purely on $D$. In this
section we consider the simplest possible such term which is
linear in $D$. The issue of uniqueness of this modification and
the results of other possible modifications will be discussed in
the next section.

By adding a linear in $D$ term to the action (\ref{opf}) we obtain
\begin{equation}
S_{op}[D,X]=Tr \Big\{ \frac{1}{2}[D,X]^2 - V(X)-iD    \Big\}.
\label{opfql}
\end{equation}
Here the coefficient in front of $D$ is taken to be imaginary to
provide hermiticity of the extra term. Being interpreted as a
derivative operator $D$ has to be anti-hermitian. Because we don't
have any Hilbert space structure yet the hermiticity relations are
to be understood as formal *-relations.

By variation of the action (\ref{opfql}) we obtain the following
modified equations of motion
\begin{equation}
[D,[D,X]] + \partial_X V(X)=0, \label{emq1l}
\end{equation}
\begin{equation}
[[D,X],X]=iI, \label{emq2l}
\end{equation}
where $I$ is the identity operator. The equations (\ref{em1}) and
(\ref{emq1l}) are the same, eqs. \ref{em2}) and (\ref{emq2l})
differ by them of $iI$ in the r.h.s. We can now solve eq.
(\ref{emq2l}) for $D$ in the form (\ref{patt}). The result
satisfying anti-hermiticity condition have the form
\begin{equation}
D=\frac{i}{2}\frac{d^2}{dX^2}-d_1(X)\frac{d}{dX}-\frac{1}{2}\partial_X
(d_1(X))-i d_0(X). \label{sold}
\end{equation}
Here $d_0$ and $d_1$ are arbitrary real functions of $x$. To find
them we should substitute $D$ in the form (\ref{sold}) to
eq.(\ref{emq1l}), which will result in the following equation
\begin{equation}
V(X)=d_0(X)+\frac{1}{2}d_1(X)^2+c. \label{vd0}
\end{equation}
Here $c$ is an arbitrary constant. Equation (\ref{vd0}) can be
solved with respect to $d_0$ and substituted to (\ref{sold}). The
resulting expression for $D$
\begin{equation}
D=\frac{i}{2}\frac{d^2}{dX^2}-d_1(X)\frac{d}{dX}-\frac{1}{2}\partial_X
(d_1(X))-\frac{i}{2}d_1(X)^2- i V(X)+ic \label{sold1}
\end{equation}
depends on an arbitrary function $d_1(X)$ and is not therefore
completely determined. Here we can recall that the equations
(\ref{emq1l},\ref{emq2l}) are invariant with respect to unitary
transformations
\begin{equation}
X\rightarrow UXU^{-1}, \ \ \ D \rightarrow UDU^{-1}. \label{untr0}
\end{equation}
Now consider the following unitary transformation
\begin{equation}
U=\exp{i \int\limits_{X_0}^{X} d_1(X)dX}. \label{untr}
\end{equation}
It is obvious that $X$ is not changed by this transformation
\begin{equation}
X'=UXU^{-1}=X.
\end{equation}
By direct computation one can check that
\begin{equation}
D'=UDU^{-1}=\frac{i}{2}\frac{d^2}{dX^2}- i V(X)+ic. \label{sgeq1}
\end{equation}
Thus the dependence on $d_1(X)$ can be removed from the final
expression for $D$ by a unitary transformation and the result is
precisely the relation between the evolution generator and, the
hamiltonian of ordinary quantum mechanics. It may be interpreted
as the Schoedinger equation when represented on states or the
Heisinberg equation when applied to operators. The only remaining
freedom is the constant $c$ indicating the arbitrariness in
defining the ground state energy of the system.

 The origin of quantum mechanical relationships between operators
 become more explicit if we consider the hamiltonian version of the
action (\ref{opf})
\begin{equation}
S_{op,h}[D,X,P]=Tr \Big\{ P[D,X]-\frac{1}{2}P^2 - V(X)    \Big\},
\label{opfh}
\end{equation}
which results in the following equivalent to (\ref{em1},\ref{em2})
equations of motion
\begin{equation}
[D,X]-P=0, \label{emh1}
\end{equation}
\begin{equation}
[D,P] + \partial_X V(X)=0, \label{emh2}
\end{equation}
\begin{equation}
[X,P]=0. \label{emh3}
\end{equation}
Now if we again add an extra term of $iD$ to it
\begin{equation}
S_{op,q}[D,X,P]=Tr \Big\{ P[D,X]-\frac{1}{2}P^2 - V(X)+iD \Big\},
\label{opfq}
\end{equation}
the modified equations of motion take the form
\begin{equation}
[D,X]-P=0, \label{emq1}
\end{equation}
\begin{equation}
[D,P] + \partial_X V(X)=0, \label{emq2}
\end{equation}
\begin{equation}
[P,X]=iI. \label{emq3}
\end{equation}
Eqs. (\ref{emq1},\ref{emq2}) are those of ordinary hamiltonian
mechanics which hold both in classical and quantum regime.
Eq.(\ref{emq3}) is the usual quantum mechanical commutation
relation. This is where the quantization comes from. The origin of
quantum mechanical commutation relation (\ref{emq3}) in this model
is analogous to that in generalized quantum mechanics proposed by
Adler \cite{adlerbook}. There a matrix extension of ordinary
mechanics of the form (\ref{higgs}) was considered. Then the
invariance with respect to unitary transformations was interpreted
as a gauge symmetry. The commutation relation of the form
(\ref{emq3}) was then induced by a term linear in the
corresponding gauge field in the action. The relation between
Adler's formulation of quantum mechanics and BFSS matrix model was
also studied by Minic \cite{minic}.

Now from (\ref{emq1},\ref{emq2},\ref{emq3}) one can derive the
equation analogous to (\ref{sgeq1}). First, from the equations
(\ref{emq2}) and (\ref{emq1}) and also (\ref{emq3}) it follows
that
\begin{equation}
[\frac{P^2}{2}+V(X),D]=0, \label{enco}
\end{equation}
which implies that
\begin{equation}
\frac{P^2}{2}+V(X)=f(D).
\end{equation}
By taking the commutator of the last equation with $X$ and using
the equations (\ref{emq3}) and (\ref{emq1}) we find that
$f(D)=iD+c$, where $c$ is an arbitrary constant playing the same
role as that in (\ref{sgeq1}), and therefore
\begin{equation}
\frac{P^2}{2}+V(X)+c=iD. \label{sgeq2}
\end{equation}
This equation coincides with (\ref{sgeq1}) given that the momentum
is represented by derivative operator $P=i\frac{d}{dX}$. Together
with the equation (\ref{emq3}) it forms the compete set of
equations of quantum mechanics of the system under consideration,
the equations (\ref{emq1}) and (\ref{emq2}) can be derived from
them.

\section{Deformations of higher order. Modified uncertainty principle and Schrodinger equation  }
\label{sec4}

 One can note that the addition of a term of $iD$ to
the action (\ref{opfq}) is not the only possible modification of
the action (\ref{opfh}) preserving the classical part of it. In
principle we can add to the action (\ref{opfh}) an arbitrary
function of $D$, $f(D)$. The function of $D$ is to be understood
as a series in powers of D.
\begin{equation}
f(D)=f_0 +f_1 D +f_2 D^2 + ... \label{dexp}
\end{equation}
As to whether this series can be understood as an asymptotic
expansion we can note that $D$ has a dimension of inverse time.
Therefore for dimension to mach we will need to introduce a
constant of the dimension of time. The expansion will be in powers
of $\tau D$ where $\tau$ is a certain fixed time parameter.
Therefore the expansion (\ref{dexp}) is asymptotic if the time
scale of phenomena considered is much larger than $\tau$. This is
satisfied if $\tau$ is Plank time, the only known fundamental time
scale in nature. Therefore it makes sense to consider the
modification of the action (\ref{opfh}) to the next (second) order
in $D$:
\begin{equation}
S_{op,q}[D,X,P]=Tr \Big\{ P[D,X]-\frac{1}{2}P^2 - V(X)+iD+\tau D^2
\Big\}. \label{opf2}
\end{equation}
For simplicity we consider the system which is classically a
harmonic oscillator, i.e. $V(X)=\omega^2 X^2$. Then the equations
of motion obtained by variation of the action (\ref{opf2}) read
\begin{equation}
[D,X]-P=0, \label{em21}
\end{equation}
\begin{equation}
[D,P] + \omega^2 X=0, \label{em22}
\end{equation}
\begin{equation}
[P,X]=iI+2\tau D. \label{em23}
\end{equation}
One can check by direct computation that the equation (\ref{enco})
holds also in this case and because $V(X)$ is quadratic this
follows from equations (\ref{em21}) and (\ref{em22}) only. Thus,
as before we have the following relation
\begin{equation}
\frac{P^2}{2}+\omega^2 X^2=f(D). \label{enco2}
\end{equation}
Finally, $f(D)$ can be found by commuting the last relation with
$X$ or $P$ using the equations (\ref{em21},\ref{em22},\ref{em23}).
We find that $f(D)=iD+\tau D^2$. Therefore instead of the
evolution equation (\ref{sgeq2}) we will have
\begin{equation}
\frac{P^2}{2}+\omega^2 X^2=iD+\tau D^2. \label{schreq2}
\end{equation}
This is a quadratic equation for $D$ and the solution is
\begin{eqnarray}
iD=\frac{1 \pm \sqrt{1-4\tau \Big[ \frac{P^2}{2}+\omega^2
X^2\Big]}}{2 \tau}
\end{eqnarray}
Now we can take the solution which has the right limit as $\tau
\rightarrow 0$ and expand it in powers of $\tau$. We will have
\begin{eqnarray}
iD= \frac{P^2}{2}+\omega^2 X^2 -\tau \Big[ \frac{P^2}{2}+\omega^2
X^2\Big]^2 + ... \label{nsheq}
\end{eqnarray}
This is what the evolution generator looks like to the first order
in $\tau$.

The commutation relation between the position and the momentum
operators is also modified. By substituting (\ref{nsheq}) into
(\ref{em23}) we find to the first order in $\tau$
\begin{equation}
[P,X]=i\bigg[I+2\tau \Big( \frac{P^2}{2}+\omega^2 X^2\Big)\bigg].
\label{mcr}
\end{equation}
Such a commutation relation is a natural generalization of the
basic commutation relations of quantum mechanics which may
underlie an uncertainty principle with minimum length and minimum
momenta or, in the case of free particle $(\omega=0)$, minimum
length only. Such possibility was considered by Kempf {\it et al}
\cite{Kempf95}, \cite{Kempf97}. For (\ref{mcr}) to be minimum
length uncertainty relation it is necessary that $\tau$ be
positive. For negative $\tau$ (\ref{mcr}) doesn't have sensible
interpretation.

The modification of the action considered in this section is
possible only if our model don't have to be general covariant. If
we consider a possibility to quantize reparameterization invariant
model then the modification of the action  (\ref{opf}) shouldn't
violate this invariance. As $Tr$ is interpreted as an integral
over $dt$ and $D$ is interpreted as the time derivative the only
possible reparameterization invariant modification of the action
(\ref{opf}) is linear in $D$. Also we can notice that the
correction to the ordinary quantum mechanical commutation relation
in (\ref{mcr}) is proportional to the hamiltonian. But in
reparameterization invariant system hamiltonian vanishes. Thus,
for general covariant systems the modification of the action
(\ref{opf}) resulting in ordinary quantum mechanics is unique.

\section{Generalization to the systems with many degrees of
freedom} \label{sec5}

 The generalization of all the above to systems
with $N$ degrees of freedom is not straightforward. If we simply
took one copy of the action (\ref{opfq}) per one degree or freedom
and considered a sum of them adding also some terms representing
interaction between these degrees of freedom then instead of
having one copy of the equation (\ref{emq3}) per each degree of
freedom we would have a single equation
\begin{equation}
\sum\limits_{i=1}^{N} [P_i,X_i]=iN, \label{wrongeq}
\end{equation}
for the whole system. This is not enough to recover the the
complete set of quantum mechanical equations of the system
considered. The same problem also arose in \cite{adlerbook}. A
solution to this problem based on thermodynamical considerations
has been given by Adler and Millard \cite{adlermillard}. It is
based on the fact that the net non-commutativity for all degrees
of freedom given by (\ref{wrongeq}) is distributed uniformly
between different degrees of freedom in thermodynamical
approximation. This is analogous to how each degree of freedom
carry a $(1/2)kT$ portion of energy at equilibrium.

Here we will give another solution to this problem which is very
natural in the context of matrix models and which is exact, i.e.
valid also apart from thermodynamical equilibrium. It is based on
a specific way of encoding of the degrees of freedom of the theory
which is extensively used in various matrix models
\cite{BFSS,IKKT,cmm}.

 Let our system have N degrees of freedom. They can be represented
 by $N$ infinite dimensional matrices commuting with each other
$X_1,X_2,...,X_N$, $[X_I,X_J]=0$. Then we can construct the
following operator
\begin{eqnarray}
{\cal X}= \left(
\begin{array}{cccc}
X_1 & 0 & \cdots & 0 \\
0 & X_2 & \cdots & \vdots \\
\vdots & \vdots & \ddots & 0 \\
0 & \cdots & 0 & X_N
\end{array} \right)=diag(X_1,X_2,..,X_N).
\label{bigx}
\end{eqnarray}
This is an infinite dimensional matrix divided into $N$ infinite
dimensional blocks located at diagonal. It acts on extended space
which is direct sum of $N$ copies of the space where a matrix
representing a single degree of freedom act.

 The matrix (\ref{bigx}) will be
treated as a whole i.e. as a single matrix-valued degree of
freedom.  Therefore we need to introduce some tools which would
allow us to identify the degrees of freedom of our interest within
this matrix. The natural additional structure we have here is that
of linear space $\mathbb{R}^N$ the dimensions of which are
associate with the degrees of freedom of the system. One can
construct linear transformations on this space. They are $N \times
N$ matrices whose entries are arbitrary $c$-numbers. Among these
matrices we can find projectors and by applying them extract the
degrees of freedom of our interest. Invertible matrices form a
group $GL(N)$. In the case of identical non-interacting particles
this is a symmetry of the theory. However, in general case this
symmetry is broken by interaction.


Now one needs to define the time evolution generator $D$ acting on
the extended space. In the present paper we will consider the
situation in which time is unique for all degrees of freedom. This
is the way how ordinary mechanics is usually described. However
keeping in mind the possibility to apply this scheme to background
independent matrix models one may notice that this framework
admits a natural generalization for multi-fingered time.

Because time is unique for all degrees of freedom $D$ must be a
$c$-number in $\mathbb{R}^N$, i.e. it must commute with all
matrices from $GL(N)$,
\begin{eqnarray}
{\cal D}= D\otimes I_{N\times N}=\left(
\begin{array}{cccc}
D & 0 & \cdots & 0 \\
0 & D & \cdots & \vdots \\
\vdots & \vdots & \ddots & 0 \\
0 & \cdots & 0 & D
\end{array} \right).
\label{bigd}
\end{eqnarray}
This is where the information about the number of degrees of
freedom of the system is encoded. The system has $N$ degrees of
freedom if there is a unitary transformation by which the
evolution generator ${\cal D}$ can be brought in the form
(\ref{bigd}) with $N$ identical blocks $D$ on the diagonal.
Transformations from $GL(N)$ keep the form (\ref{bigd}) unchanged.
Now these transformations can be used to bring an arbitrary matrix
${\cal X}$ to block-diagonal form (\ref{bigx}). Given the usual
relation between coordinate and momentum ${\cal P}=[{\cal D},{\cal
X}]$ and the form (\ref{bigd}) of the operator ${\cal D}$
canonical coordinate and momentum can be block-diagonalized
simultaneously. So the canonical momentum can be also represented
in the form ${\cal P}=diag(P_1,P_2,..,P_N)$.

Thus, the generalization of the equation (\ref{opfq}) for a system
with $N$ degrees of freedom has the following form.
\begin{equation}
S_{op,q}[D,{\cal X},{\cal P}]=Tr \Big\{ {\cal P}[D,{\cal
X}]-\frac{1}{2}{\cal P}^2 -{\cal V}[{\cal X}]+iD \Big\},
\label{opfqn}
\end{equation}
where the interaction is introduced via
\begin{eqnarray}
{\cal V}[{\cal X}]=\sum\limits_{m=1}^{\infty}{\cal X}V_{m,1} {\cal
X}V_{m,2}...{\cal X}V_{m,m}. \label{inter}
\end{eqnarray}
Here $V_{i,j}$ are fixed $N \times N$ matrices whose entries are
$c$-numbers. A model can be specified by a particular choice of
the matrices $V_{i,j}$. Specifically, if all the $V_{i,j}$ are
diagonal in the same basis where ${\cal X}$ and ${\cal P}$ are
block diagonal they introduce self interaction of the degrees of
freedom $X_I$. Interaction between $X_I$ and $X_J$, $I\not=J$, can
be introduced via non-diagonal elements of $V_{i,j}$. Note that
${\cal V}[{\cal X}]$ is a matrix of potential functions, not a
single potential function. This is a generalization of ordinary
mechanical system in which the forces may be non-conservative. In
the present paper we are interested in reproducing ordinary
mechanical systems, i.e. those with conservative forces induced by
a single overall potential $V(X_1,X_2,...,X_N)$. So we will
restrict ourselves to the situation when the matrices $V_{i,j}$
are such that ${\cal V}[{\cal X}]=V(X_1,X_2,...,X_N)\otimes
I_{N\times N}$.

 The action
(\ref{opfqn}) results in the following equations of motion
\begin{equation}
[{\cal D},{\cal X}]-{\cal P}=0, \label{emqn1}
\end{equation}
\begin{equation}
[{\cal D},{\cal P}] + \partial_{\cal X} {\cal V}({\cal X})=0,
\label{emqn2}
\end{equation}
\begin{equation}
[{\cal P},{\cal X}]=iI. \label{emqn3}
\end{equation}
which completely coincide to the equations
(\ref{emq1},\ref{emq2},\ref{emq3}) of the system with one degree
of freedom. As we mentioned above the fact that this system
describes $N$ degrees of freedom follows from the assumption that
the evolution generator ${\cal D}$ can be cast in the
block-diagonal form (\ref{bigd}) by a unitary transformation.

And there is still a question: is the set of equations
(\ref{emqn1},\ref{emqn2},\ref{emqn3}) sufficient to completely
describe quantum theory of the system? Note that the operator
commutation relations are encoded in the equation (\ref{emqn3})
which also can be rewritten as
\begin{equation}
[P_J,X_J]=iI, \label{cr1}
\end{equation}
while in canonical quantization the following set of commutation
relations is imposed
\begin{equation}
[P_J,X_K]=iI\delta_{JK}. \label{cr2}
\end{equation}
So the commutation relations (\ref{cr2}) with $J\not=K$ are missed
in the equation (\ref{cr2}). And the question is whether they can
be recovered somehow from the equations
(\ref{emqn1},\ref{emqn2},\ref{emqn3}).

We will address this question by considering a system with two
degrees of freedom. The generalization of this scheme to arbitrary
number of degrees of freedom is straightforward but require longer
calculations. Let us start with Lagrangian equivalent of
eqs.(\ref{emqn1},\ref{emqn2},\ref{emqn3})
\begin{equation}
[{\cal D},[{\cal D},{\cal X}]] + \partial_{\cal X} {\cal V}({\cal
X})=0, \label{emqnl1}
\end{equation}
\begin{equation}
[[{\cal D},{\cal X}],{\cal X}]=iI. \label{emqnl2}
\end{equation}
By taking into account that ${\cal D}=D\otimes I_{2\times 2}$ and
${\cal V}=V\otimes I_{2\times 2}$ and taking ${\cal X}$ in
block-diagonal form (\ref{bigx}) for $N=2$ the equations
(\ref{emqnl1},\ref{emqnl2}) can be rewritten as the following set
of equations for $D$
\begin{displaymath}
[ D, [ D,X_1 ] ] + \partial_{X_1} V(X_1,X_2)=0
\end{displaymath}
\begin{equation}
[ D, [ D,X_2 ] ] + \partial_{X_2} V(X_1,X_2)=0 \label{ep1}
\end{equation}

\begin{displaymath}
 [ [D,X_1 ],X_1 ]=iI
\end{displaymath}
\begin{equation}
 [ [D,X_2 ],X_2 ]=iI \label{ep2}
\end{equation}
To solve these equations for $D$ we will use a pattern similar to
(\ref{patt}). Generalized to the case of two degrees of freedom it
will have the following form
\begin{equation}
D= \sum\limits_{i,j=0}^{\infty} d_{ij}(X_1,X_2)
\left(\frac{\partial}{\partial X_1}\right)^i
\left(\frac{\partial}{\partial X_2}\right)^j. \label{patt2}
\end{equation}
The coefficients $d_{ij}$ are to be found from eqs.
(\ref{ep1},\ref{ep2}). By substituting (\ref{patt2}) into
(\ref{ep2}) we find (for convenience we introduced the following
notations for nonzero coefficients $f=d_{11}$, $g_1=d_{10}$,
$g_2=d_{01}$, $h=d_{00}$)
\begin{equation}
iD=-\frac{1}{2}\frac{\partial^2}{\partial X_1^2}-
\frac{1}{2}\frac{\partial^2}{\partial X_1^2}+
f\frac{\partial^2}{\partial X_1 X_2 }+ g_1\frac{\partial}{\partial
X_1}+ g_2\frac{\partial}{\partial X_2}+h. \label{sold2}
\end{equation}
From (\ref{sold2}) one can derive expressions for momenta
\begin{displaymath}
P_1=i[D,X_1]=-\frac{\partial}{\partial  X_1}
+f\frac{\partial}{\partial  X_2}+g_1
\end{displaymath}
\begin{equation}
P_2=i[D,X_2]=-\frac{\partial}{\partial  X_2}
+f\frac{\partial}{\partial  X_1}+g_2. \label{momenta2}
\end{equation}
It is easy to see that in this special case the commutation
relations (\ref{cr2}) with $J\not=K$ which do not enter the
equations of motion explicitly are determined by the function $f$
in (\ref{sold2}). And the question is now whether $f$ can be
determined by eqs. (\ref{em2}). To recover ordinary quantum
mechanics we need to prove that $f=0$.

Eqs. (\ref{em2}) after substituting (\ref{sold2}) and
(\ref{momenta2}) into them can be solved by considering the
resulting expression as a polynomial in derivatives and requiring
that the coefficient in front of each term vanish. The terms with
second order derivatives vanish if
\begin{equation}
\partial_{X_2}f + f\partial_{X_1}f=0 \ \ \mbox{and}  \ \
\partial_{X_1}f + f\partial_{X_2}f=0,
\end{equation}
which immediately implies that $f=const$. The later also
simplifies the condition of vanishing of the first order terms,
which is now
\begin{equation}
\partial_{X_1}g_2-f\partial_{X_2}g_2+f\partial_{X_1}g_1-\partial_{X_2}g_1=0.
\label{or1con}
\end{equation}

Now one can try to remove the linear in derivatives terms from the
expression (\ref{sold2}) for $D$ by a unitary transformation of
the type (\ref{untr0}) with $U=\exp (F)$. For this $F$ has to
satisfy the following equations
\begin{displaymath}
-\partial_{X_1}F+f\partial_{X_2}F+g_1=0
\end{displaymath}
\begin{equation}
-\partial_{X_2}F+f\partial_{X_1}F+g_2=0. \label{ufcond}
\end{equation}
This system of equations doesn't have a solution for arbitrary
$g_1$ and $g_2$. They have to satisfy a certain condition. By
taking derivatives of the equations (\ref{ufcond}) and combining
them we find that solvability condition for the system
(\ref{ufcond}) exactly coincide with (\ref{or1con}) and is
therefore satisfied automatically. Thus, there always exist a
unitary transformation by which the evolution generator can be
reduced to the form
\begin{equation}
iD=-\frac{1}{2}\frac{\partial^2}{\partial X_1^2}-
\frac{1}{2}\frac{\partial^2}{\partial X_1^2}+
f\frac{\partial^2}{\partial X_1 X_2 }+h, \label{sold2r}
\end{equation}
where $f$ is a constant. Finally, by substituting $D$ in this form
into (\ref{em2}) we find the following equations for $h$.
\begin{displaymath}
\partial_{X_1}h-f\partial_{X_2}h=\partial_{X_1}V
\end{displaymath}
\begin{equation}
\partial_{X_2}h-f\partial_{X_1}h=\partial_{X_2}V. \label{heq}
\end{equation}
Again we can derive the solvability condition for this system of
equations which is
\begin{equation}
f(\partial^2_{X_2}-\partial^2_{X_1})V=0. \label{hcond}
\end{equation}
For arbitrary $V$ there is only one solution to this equation,
$f=0$, which as it was mentioned above results in the standard
quantum mechanical commutation relations. Then the solution for
(\ref{heq}) is $h=V+c$ and the resulting expression for the
evolution generator is
\begin{equation}
iD=-\frac{1}{2}\frac{\partial^2}{\partial X_1^2}-
\frac{1}{2}\frac{\partial^2}{\partial X_2^2}+V+c. \label{sold2f}
\end{equation}
Thus, although the equations of motion derived from the action
principle (\ref{opfqn}) do not include all the equations of
quantum mechanics explicitly, they do unambiguously describe the
standard quantum mechanics.

There is however an exception. The equation (\ref{hcond}) may have
also solutions with $f\not=0$. This happens when
\begin{equation}
(\partial^2_{X_2}-\partial^2_{X_1})V=0,
\end{equation}
i.e. when the degrees of freedom are identical and the interaction
between them is linear. This is however a very special case and
it's difficult to judge on this basis whether the solutions with
$f\not=0$ lead to any new physics.

\section{Discussion}
\label{sec6} In this section we will discuss the possibility to
apply the above scheme to systems with infinite number of degrees
of freedom, i.e. to second quantized theories.

\subsection*{Field theory}
It is straightforward to write down an action analogous to
(\ref{opf}) for a scalar field. It will have the form of early
reduced models \cite{eguchikawai}

\begin{equation}
S=Tr \Big\{ [D_\mu,\Phi]^2 - V(\Phi) \Big\}, \label{fieldmm}
\end{equation}
where $Tr$ stays for $\int d^4 x$ and $D_\mu$ for
$\frac{\partial}{\partial x_\mu}$. Contrary to \cite{eguchikawai}
$D_\mu$ are now not fixed and are treated as independent
variables. It is easy to derive the equations of motion for $\Phi$
and $D_\mu$ by varying (\ref{fieldmm}) and see that the ordinary
classical scalar field theory solves those equations.

One can consider a modification of the action (\ref{fieldmm})
analogous to (\ref{opfq}) by adding a term to $S$ which depends
only on $D_\mu$. There are many possible expressions that could be
made of $D_\mu$. Here we will mention only those which are
generally-covariant.

There is no covariant term linear in $D_\mu$ that could be added
to (\ref{fieldmm}). There is a covariant first order derivative
operator acting on spinors which is the Dirac operator
$\mathord{\not\mathrel{D}}=\gamma^\mu D_\mu$. However
$\mathord{\not\mathrel{D}}$ is traceless and therefore doesn't
result in a nontrivial contribution to (\ref{fieldmm}). Thus, the
only way to get a covariant term linear in $D_\mu$ is to introduce
a fixed background operator and combine it with $D_\mu$ in such a
way that the resulting expression would be covariant. We consider
an expression depending on $D_0$ only. In this case the equations
for $D_i$ are the same as for classical system. We can solve them
in the same way thereby introducing a three dimensional space. We
can use differential structure of this space for further
construction.  The expression linear in $D_0$ can be made
covariant by multiplying it by a three dimensional density. To be
non-dynamical this density should be made completely of
coordinates. The only such density known is the three dimensional
delta function $\delta^3(x_i-x'_i)$. Thus, the expression to be
added to (\ref{fieldmm}) is
\begin{equation}
S_q=S+i Tr \sum\limits_{x'_i}^{} D_0 \delta^3 (x_i-x'_i),
\label{fieldmmq}
\end{equation}
where sum is taken over the continuum of the points of the three
dimensional manifold.

The matrix model (\ref{fieldmmq}) although very singular and
ill-defined does have the standard quantum field theory as its
solution. The question whether this solution is unique is
difficult to answer. The most problematic point is to check
whether the equation analogous to Eq. (\ref{cr2}) for $J\not=K$
holds for this model. Now it has the following involving a
continuous parameter form
\begin{equation}
[[D_0,\Phi],e^{x^iD_i}\Phi e^{-x^iD_i}]=\delta^3(x_i).
\label{cr2ft}
\end{equation}
To see whether it is a consequence of the equations of motion of
(\ref{fieldmmq}) we should solve the later by using a pattern
analogous to (\ref{patt2}) which will now involve variational
derivatives. The problem is that the series in this pattern can
not be made finite as it was done in finite dimensional case.
Therefore it is a difficult technical problem to check whether
this model has solutions other than the standard quantum field
theory.

We can consider higher order in $D_\mu$ corrections to the action
(\ref{fieldmm}) not involving metrics. The most obvious one is
\begin{equation}
S'=S+\epsilon^{\mu \nu \alpha \beta} Tr D_\mu D_\nu D_\alpha
D_\beta. \label{cst}
\end{equation}
This correction is identically zero in four dimensional space
however it is non-zero in odd dimension and can be interpreted as
Chern-Simons term.

Another possibility involves the Dirac operator
\begin{equation}
S'=S+Tr\mathord{\not\mathrel{D}}^2. \label{sp}
\end{equation}
Such term is used in the spectral action principle in
non-commutative geometry \cite{spacpr,connes} and contains in
particular Einstein-Hilbert action.

This are all the possible modifications of the action
(\ref{fieldmm}) not involving metrics.

\subsection*{Various matrix models}
If we try to apply the result of this paper for example directly
to IKKT model(\ref{sikkt}) we encounter the same problem as with
mechanical system with many degrees of freedom. The equation of
motion for $A_0$ has the form
\begin{equation}
\sum\limits_j [E_j,A_j] = iI. \label{we2}
\end{equation}
The presence of a sum over $j$ in this equation means that the
quantum commutation relations are not completely determined.

The solution to this problem would be the packing of all the
matrices $A_j$ into a single matrix. The resulting matrix model
wouldn't even have a preexisting dimension. The IKKT model then
could be derived from this model by a symmetry breaking.

Such models already exist in literature. These are cubic matrix
models \cite{cmm} based on the following action principle
\begin{equation}
S=\epsilon^{\mu \nu \alpha} Tr A_\mu A_\nu A_\alpha.
\end{equation}
This action is explicitly background independent and may be
relevant not only to string theory but also to loop quantum
gravity given the relation between loop quantum gravity and
topological field theory \cite{llt}. As this action doesn't
include any background metric no equations of the type (\ref{we2})
can appear as the sum in (\ref{we2}) is the contraction of indices
by a metric. Therefore there will be no ambiguity in definition of
quantum commutation relations derived from the action
\begin{equation}
S'=S+iTrA_0. \label{csm}
\end{equation}
To make the action (\ref{csm}) covariant we can rewrite it in the
following form
\begin{equation}
S'=S+i\epsilon^{\mu \nu \alpha}Tr\theta_{\mu \nu}A_\alpha,
\label{csm1}
\end{equation}
where all $\theta$'s are fixed. It is interesting to note that
exactly the action (\ref{csm1}) was considered as giving rise to
the space non-commutativity in non-commutative Chern-Simons theory
also playing important role in the description of quantum Hall
effect \cite{noncomcs}. This may imply a possible relation between
space-time non-commutativity in non-commutative geometry and
quantum mechanical non-commutativity. This issue will be discussed
elsewhere.

\subsection*{A comment on diffeomorphism invariance}
The mechanical model considered in this paper was not
diffeomorphism invariant. It is worth making some comments on how
to apply this formalism to a diffeomorphism invariant theory. As
we mentioned above extra terms in the action resulting in
quantization do not violate diffeomorphism invariance while almost
any other extra term would violate it. It is suggestive that if we
require diffeomorphism invariance we reproduce the known physics.
Now the question is how to cast the ordinary diffeomorphism
invariance of a field theory in the language of matrix models.

When we consider an action of a field theory we may notice that
there are generally two kinds of entries in it. First, it contains
fields. This may be fields describing geometry of the manifold
such as metrics and connections as well as  matter fields. All the
fields can be evaluated as functions of coordinates on the
manifold $x^{\mu}$. The other type of entries of the classical
action are derivative operators $\partial_\mu$. We generally
assume that these entries form the following Poisson algebra with
respect to commutator:
\begin{equation}
[x^\mu,x^\nu]=0, \label{pa1}
\end{equation}
\begin{equation}
[\partial_\mu,\partial_\nu]=0, \label{pa2}
\end{equation}
\begin{equation}
[\partial_\mu,x^\nu]=\delta_\mu^\nu. \label{pa3}
\end{equation}

It is natural to identify the action of diffeomorphism group with
the coordinate transformation, ${x^\mu}'={x^\mu}'(x^\mu)$. Below
for simplicity we will consider small transformations,
\begin{equation}
{x^\mu}'=x^\mu+f^\mu(x^\mu), \label{sdiff1}
\end{equation}
where $f^\mu \ll x^\mu$. By analogy with hamiltonian mechanics we
can think of $x^\mu$ as canonical coordinates parameterizing
configuration space and of $\partial_\mu$ as canonical momenta,
with (\ref{pa1},\ref{pa2},\ref{pa3}) being the Poisson bracket
relations. For any transformation of configuration space  of the
form (\ref{sdiff1}) there exists a canonical transformation of the
phase space with the generating function
\begin{equation}
D=f^\mu(x^\mu)\partial_\mu \label{diffgen}
\end{equation}
 defined by
\begin{equation}
{x^\mu}'=x^\mu+[D,x^\mu]=x^\mu+f^\mu(x^\mu), \label{ctc}
\end{equation}
\begin{equation}
\partial_\mu'=\partial_\mu+[D,\partial_\mu]=\partial_\mu+\frac{\partial f^{\nu}}{\partial
x^{\mu}}\partial_\nu. \label{ctm}
\end{equation}
One can say that under the action of the diffeomorphism group the
algebraic elements $x^\mu$ and $p_\mu$ transform according to
(\ref{ctc},\ref{ctm}). This action preserves the relations
(\ref{pa1},\ref{pa2},\ref{pa3}). If we represent $x^\mu$ and
$p_\mu$ by hermitian operators on certain Hilbert space $H$
diffeomorphisms (\ref{ctc},\ref{ctm}) will be represented by
unitary transformations in this space. Therefore the
diffeomorphism group is a subgroup of the group of all unitary
transformations in $H$.

However the diffeomorphism transformation introduced above are not
those playing the crucial role in General Relativity. Here one has
to recall that there are two kinds of diffeomorphisms: active and
passive. Any field theory is trivially invariant with respect to
passive diffeomorphisms which act simultaneously on coordinates
and fields. On the other hand only general relativistic theory is
invariant with respect to active diffeomorphisms which act on
fields only. For detailed discussion on the distinction between
active and passive diffeomorphisms see \cite{RovelliGaul}.

In order to distinguish active and passive diffeomorphisms here
one needs to see how they act on the fields. For simplicity let us
take all the fields in the theory to be scalars $\phi_n (x^\mu)$,
$n=1,...,N$. If $\phi_n$ is a smooth function of $x^\mu$ then
given the transformation (\ref{ctc}) for $x^\mu$ one can see that
$\phi_n (x^\mu)$ transforms with respect to diffeomorphisms as
follows
\begin{equation}
{\phi_n}' (x^\mu)=\phi_n (x^\mu)+[D,\phi_n (x^\mu)]=\phi_n
(x^\mu)+\frac{\partial \phi_n (x^\mu)}{\partial
x^\nu}f^\nu(x^\mu). \label{ftr}
\end{equation}
This transformation law together with (\ref{ctm}) is exactly the
way fields and derivatives should transform with respect to
passive diffeomorphisms. Under active diffeomorphisms fields
should transform according to (\ref{ftr}) while derivative
operators should remain unchanged.

Since the action principle doesn't depend on coordinates
explicitly it is convenient to rewrite the relations
(\ref{pa1},\ref{pa2},\ref{pa3}) in terms of fields and derivatives
operators. They have the following form
\begin{equation}
[\phi_n,\phi_m]=0, \label{pa1d}
\end{equation}
\begin{equation}
[\partial_\mu,\partial_\nu]=0, \label{pa2d}
\end{equation}
\begin{equation}
[[\partial_\mu,\phi_n]\phi_m]=0. \label{pa3d}
\end{equation}
It is interesting to note that unlike the relations
(\ref{pa1},\ref{pa2},\ref{pa3}) the relations
(\ref{pa1d},\ref{pa2d},\ref{pa3d}) are invariant not only with
respect to passive but also with respect to active
diffeomorphisms, which can be checked by direct computation. Thus,
for a field theory containing scalar fields only it is easy to
include the active diffeomorphisms in this framework. They are a
subgroup of the full group of unitary transformations of the
matrix model with respect to which derivative operators are
$c$-numbers.

The situation becomes more difficult if we consider a matrix model
based on a gauge theory. Gauge fields have the following
transformation low with respect to diffeomorphisms
\begin{equation}
A_\mu'=A_\mu +f^\mu F_{\mu \nu} \not=A_\mu [D,A_\mu],
\end{equation}
where $F_{\mu \nu}$ is a curvature tensor of the connection
$A_\mu$. So if we consider a gauge field $A_\mu$ as a matrix or
equivalently consider the full covariant derivative
$D_\mu=\partial_\mu+A_\mu$ as a matrix then the action of the
diffeomorphism group is not a unitary transformation in the space
of such matrices. The situation can be fixed by multiplying the
covariant derivative by a frame field $e^\mu_a$ where the index
$a$ refers to the tangent space. Given that $e^\mu_a$ transforms
as a vector the matrices $G_a$ defined by
\begin{equation}
G_a=e^\mu_a D_\mu
\end{equation}
have the following transformation law with respect to
diffeomorphisms
\begin{equation}
G_a'=G_a+[f^\mu D_\mu, G_a]=G_a+[D,G_a].
\end{equation}
So the active diffeomorphisms are unitary transformations in the
space of matrices $G_a$. But one can easily see that the passive
diffeomorphisms have exactly the same action on $G_a$ which is the
property of not only generally relativistic theories . Therefore
the question of the meaning of diffeomorphism invariance in gauge
theory matrix models is very subtle.

Finnaly, it is worth making a comment on how the relations of the
type (\ref{pa1d},\ref{pa2d},\ref{pa3d}) are understood in this
framework. If the algebra ${\mathcal A}=\{\phi_n,\partial\}$ is an
algebra of fields and derivative operators on a certain classical
differential manifold then the relations
(\ref{pa1d},\ref{pa2d},\ref{pa3d}) are satisfied.  We can invert
the statement and say that if the relations
(\ref{pa1d},\ref{pa2d},\ref{pa3d}) are satisfied than the algebra
${\mathcal A}$ admits a representation as a classical manifold.
Possible deformations of the relations
(\ref{pa1d},\ref{pa2d},\ref{pa3d}) are to be understood as
generalizations of the notion of manifold. In particular
deformations of the relation (\ref{pa1d}) can be understood in the
framework of Non-commutative Geometry \cite{connes}. Thus, matrix
models can be understood as a framework in which the relations
(\ref{pa1d},\ref{pa2d},\ref{pa3d}) (or part of them) are not fixed
in the beginning. Instead  these relations  are determined by
dynamics, i.e. derived from the equations of motion of the theory.

\section*{Acknowledgements}
I am grateful to Stephen Adler, Satoshi Iso, Achim Kempf, and
especially Lee Smolin for useful discussions.

\end{document}